\documentstyle[psfig]{europhys} 
\newcommand{\lesssim}{\raisebox{-.7ex}{\,\protect{$\stackrel{<}{\sim}$}\,}}
\newcommand{\gtrsim}{\raisebox{-.7ex}{\,\protect{$\stackrel{>}{\sim}$}\,}}

\def\etal{{\hbox{{\tenit\ et al.\/}\tenrm :\ }}}

\def\And{{\rm and\ }}

\def\stars{\bigskip\centerline{***}\medskip}

\newif\ifboo \boofalse

\def\Review#1{\boofalse{\it #1},}
\def\Name#1{{\sc #1},}
\def\Vol#1{\ifboo Vol. {\bf #1}\else{\bf #1}\fi}
\def\Year#1{\ifboo #1\else(#1)\fi}
\def\Book#1{\bootrue{\it #1},}
\def\Page#1{\ifboo {\rm p. #1}\else{\rm #1}\fi}

\begin{document}
\euro{xx}{x}{xx-xx}{xxxx}
\Date{}
\shorttitle{C. GRIMALDI \etal PAULI SUSCEPTIBILITY OF
ETC.}
\title{{\Large Pauli susceptibility of nonadiabatic Fermi liquids}} 
\author{C. Grimaldi\inst{1} and
L. Pietronero\inst{2}}
\institute{\inst{1} \'Ecole Politechnique F\'ed\'erale de Lausanne, 
DMT-IPM CH-1015 Lausanne, Switzerland\\
\inst{2} Dipartimento di Fisica and Unit\'a INFM,
Universit\'a di Roma  ``La Sapienza''
P.le A. Moro 2, I-00185 Roma}
\rec{}{}
\pacs{
\Pacs{63}{20$Kr$}{Phonon-electron and phonon-phonon interactions}
\Pacs{71}{38$+i$}{Polarons and electron-phonon interactions}
\Pacs{76}{30$-v$}{Electron paramagnetic resonance and relaxation}
}

\maketitle
\begin{abstract}
The nonadiabatic regime of the electron-phonon interaction
leads to behaviors of some physical measurable quantities
qualitatively different from those expected from the Migdal-Eliashberg
theory. Here we identify in the Pauli paramagnetic susceptibility $\chi$
one of such quantities and show that the nonadiabatic
corrections reduce $\chi$ with respect to its adiabatic limit.
We show also that the nonadiabatic regime induces an isotope
dependence of $\chi$, which in principle could be measured.

\end{abstract}

When the Fermi energy $E_F$ is anomalously small, as in
high-$T_c$ cuprates \cite{uemura}  and in the fullerene 
compounds \cite{gunna1}, the
Migdal-Eliashberg (ME) approach \cite{migdal,elia} may result inadeguate in
describing the interplay between charge carriers and
phonons. For example, the alkali-doped fullerenes (A$_3$C$_{60}$)
have Fermi energies of order $0.25$ eV \cite{gunna1} and intramolecular 
phonon modes with frequencies $\omega_0$ in the range between 
$20$ meV and $0.2$ eV \cite{hebard}. In this case, the adiabatic parameter
$\omega_0/E_F$ lies somewhere between $0.1$ and $0.9$,
depending on which phonon modes most couple to the electrons.
The main consequence is that the electron-phonon vertex corrections
may no longer be negligible, as assumed in the ME framework,
and a generalization of the theory is required to include
the nonadiabatic contributions \cite{grima1}.

This generalization 
In terms of the electron-phonon coupling $\lambda$ and
the adiabatic parameter $\omega_0/E_F$, the ME regime applies
for $\lambda \lesssim 1$ and $\omega_0/E_F\ll 1$. Therefore,
a generalization beyond the ME framework is required when
$\lambda\gtrsim 1 $ and/or $\omega_0/E_F$ is no longer negligible. 
However, when $\lambda$ is larger than some critical value $\lambda_c$ 
(which is of order one or larger), the system evolves
toward a polaronic regime characterized by strong electron-lattice
correlations. This holds true even in the adiabatic case
in which the charge carriers aquire large effective masses.
On the other hand, 
a region in the $\lambda$-$\omega_0/E_F$ plane different from the
one leading to polaronic states
is defined by $\lambda\lesssim 1$ and $\omega_0/E_F$ finite.
Within this region, where the charge carriers are weakly interacting
nonadiabatically with phonons, the nature of quasiparticles 
is different from both the ME and the polaronic ones.
In such a nonadiabatic regime we shall speak of nonadiabatic
Fermi liquids (or nonadiabatic fermions), to stress the difference 
from the ME and the polaronic limits. 
In practice, such a regime can be described by a perturbative
approach where $\lambda\omega_0/E_F$ plays the role of
the small parameter of the theory \cite{grima1,grima2}.
 Various comparisons with exact results (for the one electron case)
\cite{capone}
and quantum Monte Carlo calculations \cite{free} point toward 
the reliability of such a perturbative description.

At the zeroth order in $\lambda\omega_0/E_F$,
the nonadiabatic theory coincides with the ME limit while for finite
values of $\lambda\omega_0/E_F$ the nonadiabatic fermions
display anomalous behaviors.
In this situation, several properties are modified 
and a very important question 
regards the possibility to observe
some fingerprints of such a nonadiabatic regime. Furthermore,
in order to be considered as possible evidences, 
such fingerprints should be searched among those
physical quantities for which some well established property in the
ME regime results {\it qualitatively} modified in the nonadiabatic one.
In order to clarify this statement, let us consider for example
the electron-phonon renormalized charge carrier mass $m^*$.
In the ME regime $m^*=(1+\lambda)m$ \cite{grimvall}, where $m$ is the
bare mass and $\lambda$ is the electron-phonon coupling.
Since $\lambda$ is independent of the ion-mass \cite{carbotte}, no isotope
effect is expected for $m^*$. However, when the nonadiabatic
contributions are no longer negligible, $m^*$ aquires an ion-mass
dependence which leads to a non-zero isotope coefficient
$\alpha_{m^*}$ \cite{grima3}.
The effective mass $m^*$ represents therefore a clear example
of a quantity for which a well established property in the ME
regime ($\alpha_{m^*}=0$) is drastically modified in the
nonadiabatic one ($\alpha_{m^*}\neq 0$). So far, strong
evidences for isotope-dependent $m^*$ have been reported for
YBa$_2$Cu$_3$O$_{6+x}$ and La$_{2-x}$Sr$_x$CuO$_4$ \cite{zhao} and 
theoretical calculations have shown that already the inclusion
of the first nonadiabatic vertex correction to the ME limit 
provides values of $\alpha_{m^*}$ with sign and order of magnitude
in agreement with those estimated by the experiments \cite{grima3}.

Another property typical in the ME regime which is instead
strongly altered by the nonadiabatic contributions is the 
non-magnetic impurity dependence of the critical temperature
$T_c$ of an homogeneous $s$-wave superconductor. For a 
conventional superconductor, weak disorder does not influence
the critical temperature as stated by Anderson's theorem \cite{anderson}.
On the contrary, since the electron-phonon vertex corrections
are very sensitive to the amount of disorder, the critical temperature
of a $s$-wave nonadiabatic superconductor can be strongly 
lowered by the impurities \cite{scattoni}. 
Such a peculiar behavior is also accompained by an anomalous impurity
dependence of the isotope coefficient of $T_c$. 
So far, reduction of $T_c$ driven by disorder for $s$-wave superconductors
has been reported for K$_3$C$_{60}$ \cite{watson} and 
Nd$_{2-x}$Ce$_3$CuO$_{4-\delta}$ \cite{woods}.

In this paper we consider another measurable quantity which could be considered
as a test for the breakdown of Migdal's theorem: the Pauli
paramagnetic susceptibility $\chi$. Here, the characteristic feature
in the ME regime ($\omega_0/E_F\ll 1$) is that the electron-phonon
interaction does not renormalize the Pauli susceptibility so that
$\chi$ is independent of $\lambda$ and $\omega_0$ \cite{grimvall}. 
In the ME regime
therefore $\chi\equiv \chi_P =\mu_B^2 N(0)$, where $\mu_B$ is the Bohr
magneton and $N(0)$ is the electron density of states at the Fermi level.
In principle, therefore, a measure of $\chi$ via for example
electron paramagnetic resonace (EPR) is unaffected by the electron-phonon 
interaction and provides an estimate of the electronic density
of states $N(0)$ which however is renormalized by many-electrons 
effects (Stoner enhancement)\footnote{In the present discussion
we shall consider the many-electrons effects as being
already contained in $N(0)$.}. 

The interesting aspect of $\chi$ is that, as we show below, 
when $\omega_0/E_F$ is no longer negligible $\chi$
aquires a phonon renormalization and becomes dependent on both
$\lambda$ and $\omega_0$. This result can be of importance for
two reasons. First,
it leads to re-consider the estimates of
the electron density of states obtained by EPR measurements, since these
estimates have been based on the phonon-independent ME form of $\chi$.
Second, and more importantly,
the nonadiabatic renormalization of $\chi$ induces a non-zero
isotope effect which, in principle, could be measured.

To evaluate the Pauli susceptibility we make use of the static
limit of the Kubo formula:\cite{mahan}

\begin{equation}
\label{kubo}
\chi(T)=\lim_{{\bf q}\rightarrow 0}\mu_B^2\int^{\beta}_0 \! d\tau
\langle T_{\tau}S_z({\bf q},\tau)S_z(-{\bf q},0)\rangle,
\end{equation}
where $\beta$ is the inverse temperature $T$ and

\begin{equation}
\label{S}
S_z({\bf q})=\sum_{{\bf k},\sigma=\pm 1}\sigma c^{\dagger}_{{\bf k}+{\bf q} 
\sigma}c_{{\bf k} \sigma},
\end{equation}
where $c^{\dagger}_{{\bf k}\sigma}$ ($c_{{\bf k}\sigma}$) is the creation
(annhilication) operator for electron with momentum ${\bf k}$ and
spin direction $\sigma=\pm 1$.

In what follows, we shall focus on the evaluation of Eq.(\ref{kubo})
for a system of electrons interacting with phonons through the
coupling $g({\bf q})$. In terms of electron and phonon Green's functions,
Eq.(\ref{kubo}) reduces to the following general expression:

\begin{equation}
\label{kubo2}
\chi(T)=-\lim_{{\bf q}\rightarrow 0}\mu_B^2 T\sum_m
\sum_k G(m,{\bf k}+{\bf q})G(m,{\bf k})
\Gamma({\bf k}+{\bf q},{\bf k};m),
\end{equation}
where $\omega_m=(2m+1)\pi T$ and

\begin{equation}
\label{green}
G(m,{\bf k})=\left[i\omega_m-\epsilon({\bf k})-
\Sigma(m,{\bf k})\right]^{-1},
\end{equation}
is the Green's function for an electron with dispersion $\epsilon({\bf k})$
and electron-phonon self-energy $\Sigma(m,{\bf k})$. 
In Eq.(\ref{kubo2}),
$\Gamma({\bf k}+{\bf q},{\bf k};m)$ is the irreducible
electron-phonon vertex function which is given by all diagrams
which cannot be separated into two different parts by cutting
a single electron or phonon propagator line. The reducible
part of the vertex function gives in fact zero contribution
when the summation over the spin indeces is performed in 
eqs.(\ref{kubo}-\ref{S}) \cite{mahan}.

In this paper we compute eq.(\ref{kubo2}) by employing a self-consistent
calculation which amount to evaluate $\Sigma(m,{\bf k})$
in the non-crossing approximation. For dispersionless phonons with
frequency $\omega_0$, we consider therefore the electron-phonon self-energy 
as given by:

\begin{equation}
\label{selfME0}
\Sigma(n,{\bf k})=T\sum_{m{\bf k}'}
g({\bf k}-{\bf k}')^2
\frac{\omega_0^2}{(\omega_n-\omega_m)^2-\omega_0^2}
G(m,{\bf k}').
\end{equation}
In the above equation we have implicitly assumed that the
phonons are already renormalized and that $\omega_0$
is a dressed phonon frequency. In a conserving approach, the
vertex function resulting from the non-crossing approximation for 
$\Sigma(m,{\bf k})$ is given by all the ladder contributions.
Therefore the vertex function satisfies the following ladder
equation:

\begin{eqnarray}
\label{verladd1}
\Gamma({\bf k}+{\bf q},{\bf k};n+m,n)=1+ 
&&T\!\sum_{m'{\bf k}'}g({\bf k}-{\bf k}')^2 
\frac{\omega_0^2}{(\omega_n-\omega_{m'})^2+\omega_0^2}
G(m',{\bf k}')G(m'+m,{\bf k}'+{\bf q})
\nonumber \\
&&\times\Gamma({\bf k'}+{\bf q},{\bf k}';m'+m,m').
\end{eqnarray}
Actually, from eq.(\ref{kubo2}), to evaluate $\chi$ we only 
need to retain the static
limit of eq.(\ref{verladd1}) which is given by setting first 
$\omega_m=0$ and after ${\bf q}=0$.
As already shown in Refs.\cite{grima2}, 
if we exchange the order of the two limits,
the resulting dynamical limit of the vertex will be in general different
from the static one. Therefore setting $\omega_m=0$ and ${\bf q}=0$
in both hand sides of eq.(\ref{verladd1}) may give a non well defined 
result  because in that point the vertex in non-analytic.
However, as we shall show below, the computing procedure we employ in
handling the vertex function automatically provides the correct
static limit by simply setting $\omega_m=0$, ${\bf q}=0$ in 
eq.(\ref{verladd1}), regardless of the order of the two limits. 
Therefore, by setting $\lim_{{\bf q}\rightarrow 0}
\Gamma({\bf k}+{\bf q},{\bf k};n,n)
=\Gamma_s({\bf k},n)$, the static limit of eq.(\ref{verladd1})
reduces to:

\begin{equation}
\label{verladd2}
\Gamma_s({\bf k},n)=1+ 
T\!\sum_{m'{\bf k}'}g({\bf k}-{\bf k}')^2 
\frac{\omega_0^2}{(\omega_n-\omega_{m'})^2+\omega_0^2}
G(m',{\bf k}')^2\Gamma_s({\bf k}',m').
\end{equation}

Without loss of generality, the solution of the set of equations (\ref{green}),
(\ref{selfME0}) and (\ref{verladd2}) can be found by using a
structureless electron-phonon interaction $g({\bf q})\equiv g^2$.
The resulting self-energy is then momentum independent and for
a system with a half-filled
electron band of constant DOS over the entire bandwidth $2E_F$,
the self-energy can be written as $\Sigma(n)=i\omega_n-iW_n$,
where 

\begin{equation}
\label{selfME2}
W_n=\omega_n+\lambda\pi T\sum_m\frac{\omega_0^2}
{(\omega_n-\omega_m)^2+\omega_0^2}\frac{2}{\pi}
\arctan\left(\frac{E_F}{W_m}\right),
\end{equation}
is the renormalized electron frequency and $\lambda=g^2 N(0)$
is the electron-phonon coupling. Within the same approximation scheme,
$\Gamma_s({\bf k},n)$ becomes momentum-independent and the resulting
vertex function $\Gamma_s(n)$ satisfies the following equation:

\begin{equation}
\label{verladd3}
\Gamma_s(n)=1-\lambda T\sum_{m'}
\frac{\omega_0^2}{(\omega_n-\omega_{m'})^2+\omega_0^2} 
\frac{2E_F}{W_{m'}^2+E_F^2}\Gamma_s(m') .
\end{equation}
We can verify that the above equation gives indeed the static limit 
of the vertex by
neglecting the renormalization of the frequency, 
$W_{m'}\rightarrow\omega_{m'}$,
and by performing the zero temperature limit. In this way,
to the first order in $\lambda$ and at zero external 
frequency, Eq.(\ref{verladd3})
becomes:

\begin{equation}
\label{verladd4}
\Gamma_s(0)=1-\lambda\int\frac{d\omega}{2\pi}
\frac{\omega_0^2}{\omega^2+\omega_0^2}
\frac{2E_F}{\omega^2+E_F^2} 
=1-\lambda\frac{\omega_0}{\omega_0+E_F}.
\end{equation}
$\Gamma_s(0)$ coincides therefore with the static limit 
already calculated in the perturbation theory.\cite{grima1,grima2}

We are now in the position to evaluate the Pauli susceptibility.
Since both the self-energy and the vertex function are independent
of the momentum, equation (\ref{kubo2}) can be analitically
integrated over the energy
and the final expression for $\chi(T)$ reduces to:

\begin{equation}
\label{kubo3}
\chi(T)=\chi_P T\sum_m\frac{2E_F}{E_F^2+W_m^2}\Gamma_s (m),
\end{equation}
where $\chi_P=\mu_B^2 N(0)$, and $W_m$ and $\Gamma_s(m)$ are 
the solution of equations (\ref{selfME2}) and (\ref{verladd3}), 
respectively. 

We solve the set of equations (\ref{selfME2}), (\ref{verladd3})
and (\ref{kubo3}) for a temperature $T/\omega_0=0.02$ and
differtent values of $\lambda$ and $\omega_0/E_F$.
The frequency summations appearing both in the self-energy (\ref{selfME2}) 
and in the vertex function (\ref{verladd3}) is truncated 
at the frequency cut-off $\omega_c=(2N+1)\pi T$ with $N=400$
corresponding to $\omega_c\simeq 50\omega_0$. The solutions
of (\ref{selfME2}) and (\ref{verladd3}) are then calculated
by iteration and the results are plugged into eq.(\ref{kubo3}).
The high-frequency part ($\omega_m>\omega_c\gg\omega_0$) of the 
summation in eq.(\ref{kubo3}) is calculated by setting
$W_m=\omega_m$ and $\Gamma_s(m)=1$, since in this high-frequency
region the contribution from the electron-phonon coupling is
negligible. 
The procedure outlined above permits to estimate the zero
temperature susceptibility $\chi$ also for the smallest value 
of $\omega_0/E_F$ we used
in the calculations ($\omega_0/E_F=0.01$).

In Fig. 1 we show the zero temperature calculated
Pauli susceptibility as a function of the adiabatic parameter 
$\omega_0/E_F$ and for different values of the electron-phonon
coupling constant $\lambda$. When $\omega_0/E_F\rightarrow 0$,
$\chi$ approaches its free-electron value $\chi_P$, irrespectively of 
the value of $\lambda$ and we recover therefore the result
of the ME theory. Instead,
when $\omega_0/E_F$ is larger than zero, $\chi$ becomes
dependent of $\lambda$ and results to be always lowered with 
respect to $\chi_P$.
In Fig. 2 $\chi/\chi_P$ is plotted as a function of the electron-phonon
coupling $\lambda$ for different values of $\omega_0/E_F$.
For small values of $\omega_0/E_F$, $\chi/\chi_P$ decreases 
almost linearly with $\lambda$. 
The main result of our calculations is therefore that $\chi(0)/\chi_P<1$
as soon as $\omega_0/E_F>0$. Preliminary calculations including higher
orders vertex corrections confirm this feature.

The reduction of the Pauli susceptibility induced by the electron-phonon 
interaction when $\omega_0/E_F$ is finite requires to re-consider the
estimates of the electron density of states based on EPR measurements
\cite{wong,tanigaki}. 
In these estimates, in fact, the measured $\chi$ is fitted with the
ME expression of the susceptibility

\begin{equation}
\label{chiME}
\chi\propto N(0)\sim\frac{N_0(0)}{1-I},
\end{equation}
where in the last equality we have explicitly separated $N(0)$
into the free-electron form $N_0(0)$ and the Stoner enhancement
$1/(1-I)$ given by the many-electrons effects. Theoretical
estimations of $1/(1-I)$ permit therefore to obtain $N_0(0)$
from the experimental $\chi$ \cite{gunna2}. However, this procedure may 
sistematically underestimate $N_0(0)$ if $\omega_0/E_F$
is no longer negligible like in the fullerene compounds. In fact,
in view of the previous results,  $N_0(0)$ of eq.(\ref{chiME}) 
should be replaced by
$N_0^{*}(0)\simeq N_0(0)f(\lambda,\omega_0/E_F)$, where
the function $f$ takes into account the phonon renormalization effects
and is less than the unity. From the calculations showed in
Figs. 1 and 2, $f$ can be as small as $\sim 0.8\,-\,0.7$, 
leading to an underestimation of the bare density 
of states $N_0(0)$ of $\sim 20\,-\,30\%$.

Another remarkable feature of the nonadiabatic phonon 
renormalization is the lattice induced isotope effect on $\chi$.
From Fig.1 in fact it is obvious that a change in frequency $\omega_0$
induces a lowering of $\chi$. Such a change of $\omega_0$ can be 
induced by isotope substitution leading therefore to a non-zero 
value of the isotope coefficient:

\begin{equation}
\label{iso}
\alpha_{\chi}=-\frac{d\log\chi}{d\log M}=\frac{1}{2}
\frac{d\log\chi}{d\log(\omega_0/E_F)},
\end{equation}
where $M$ is the ion mass and, in the last equality, we have 
used $\omega_0\propto (M)^{-1/2}$
(note that in the nonadiabatic regime $\chi$ depends also on $\lambda$, 
however $\lambda$ is independent of $M$).
In Fig. 3 we show the numerical evaluation of eq.\ref{iso} as a function
of $\omega_0/E_F$ and for different values of $\lambda$. As expected, 
the resulting isotope coefficient $\alpha_{\chi}$ vanishes at the adiabatic 
limit. However, for nonzero values of $\omega_0/E_F$, it becomes negative
and for ordinary values of $\lambda$ can be of order $-0.05$. This is a rather
small value, nevertheless it provides a clear indication of nonadiabaticity.
It would be extremely interesting to investigate experimentally the presence 
or the absence of an isotope effect on $\chi$ in the fullerene compounds.
The outcome of such kind of experiment could provide us with an estimate
of $\omega_0/E_F$ and therefore of the degree of nonadiabaticity in such
narrow band materials.

\stars{}

\begin{figure}
\protect
\centerline{\psfig{figure=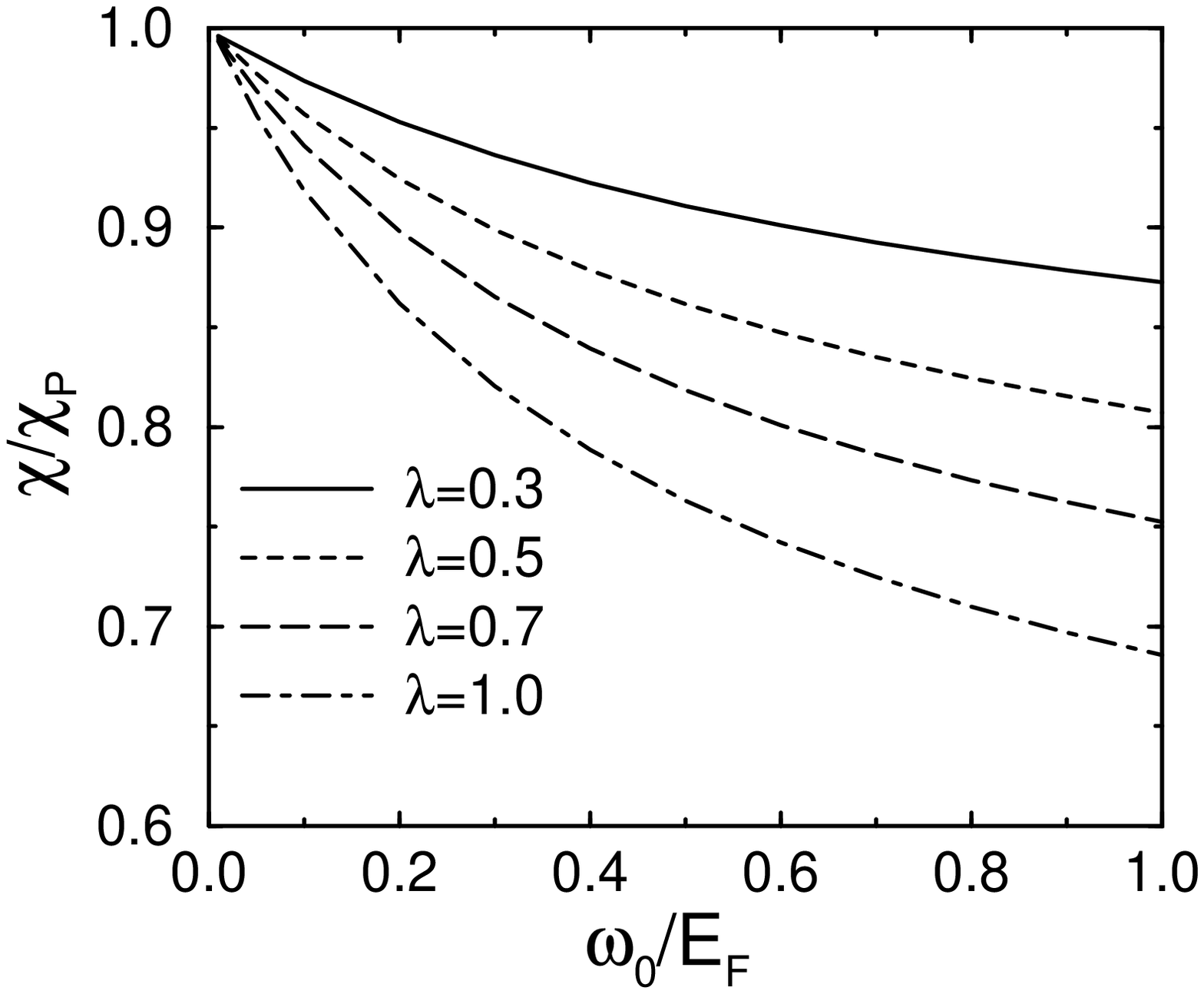,width=10cm}}
\caption{$\omega_0/E_F$ dependence of the Pauli susceptibility
$\chi$ for different values of the electron-phonon
coupling constant $\lambda$.}
\label{fig1}
\end{figure}

\begin{figure}
\protect
\centerline{\psfig{figure=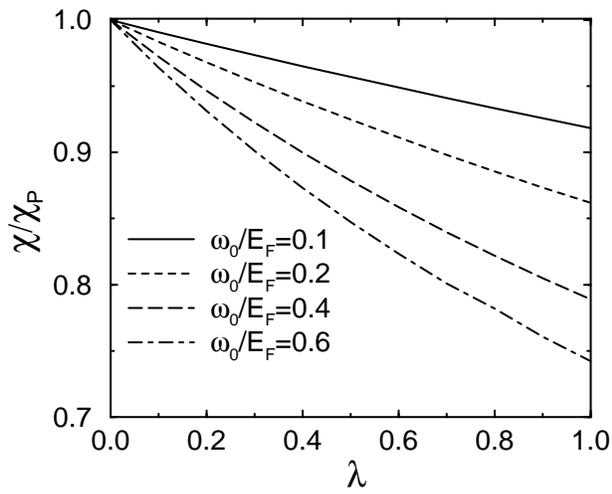,width=10cm}}
\caption{Pauli susceptibility $\chi$ as a function of the
electron-phonon coupling constant $\lambda$ for different values 
of the adiabatic parameter $\omega_0/E_F$.}
\label{fig2}
\end{figure}

\begin{figure}
\protect
\centerline{\psfig{figure=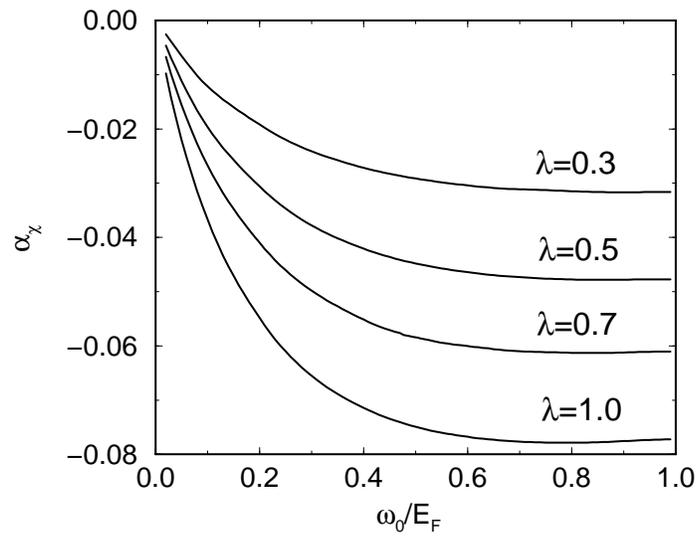,width=10cm}}
\caption{Isotope coefficient $\alpha_{\chi}$ of the
Pauli susceptibility for different values of the
electron-phonon coupling $\lambda$.}
\label{fig3}
\end{figure}

\end{document}
\bye